\UseRawInputEncoding
\documentclass[preprint]{elsarticle}

\usepackage[T1]{fontenc}
\usepackage[utf8]{inputenc}
\usepackage[english]{babel}
\usepackage{amsmath}
\usepackage{graphicx}
\usepackage{booktabs,multirow}
\usepackage{amssymb}
\usepackage{amsthm}	
\usepackage{amsfonts}
\usepackage[binary-units=true]{siunitx}
\usepackage{geometry}	

\usepackage[skins,theorems]{tcolorbox}
\tcbset{myinner/.style={no shadow,shrink tight,extrude by=1mm,colframe=blue,
  boxrule=0.4pt,frame style={opacity=0.25},interior style={opacity=0.5}}}

\usepackage{arydshln}

\usepackage{overpic}

\graphicspath{{Figures/}}

\newcommand{\rea}[1]{\mathbb{R}^{#1}}
\newcommand{\dotf}[1]{\dot{\mathbf{#1}}}
\newcommand{\dotfs}[1]{\dot{\boldsymbol{#1}}}
\newcommand{\ddotfs}[1]{\ddot{\boldsymbol{#1}}}

\newcommand{\normv}[1]{\left\vert#1\right\vert}

\newtheorem{thm}{Theorem}
\newtheorem{rem}{Remark}
\newtheorem{cor}{Corollary}

\begin{document}




\title{Unified Force and Motion Adaptive-Integral Control of \\ Flexible Robot Manipulators\tnotetext[mytitlenote]}
\tnotetext[mytitlenote]{This work was supported by the Project HOMPOT, grant number P20\_00597, under framework PAIDI 2020.}

\author[MW]{Carlos R. de Cos}
\ead{cdecos@mathworks.com}

\author[US]{Jos\'e \'Angel Acosta\corref{corauth}}
\ead{jaar@us.es}

\cortext[corauth]{Corresponding author}

\affiliation[MW]{organization={The MathWorks, AB},
city={Stockholm},
country={Sweden}}

\affiliation[US]{organization={Dept. Ingenieria de Sistemas y Automatica, Universidad de Sevilla},
city={Sevilla},
postcode={41092},
country={Spain}}

\begin{abstract}
In this paper, an adaptive nonlinear strategy for the motion and force control of  flexible manipulators is proposed. The approach provides robust motion control until contact is detected when force control is then available--without any control switch--, and vice versa. This self-tuning in mixed contact/non-contact scenarios is possible thanks to the unified formulation of force and motion control, including an integral transpose-based inverse kinematics and adaptive-update laws for the flexible manipulator link and contact stiffnesses. Global boundedness of all signals and asymptotic stability of force and position are guaranteed through Lyapunov analysis. 
The control strategy and its implementation has been validated using a low-cost basic microcontroller and a manipulator with 3 flexible joints and 4 actuators. Complete experimental results are provided in a realistic mixed contact scenario, demonstrating very-low computational demand with inexpensive force sensors.
\end{abstract}




\maketitle


\section{Introduction}

Increase interest on flexible manipulation has arisen mainly due to its potential for applications under weight constraints and physical interaction with the environment (possibly human). The resilience shown by these manipulators limits the negative effects of unforeseen impacts or other surrounding conditions, thus improving the stability and compliance under tasks involving physical interaction with the environment \cite{hogan22}.
\begin{figure}[htbp]
\centering
\includegraphics[trim = 0pt 0pt 2pt 0pt,clip,width=0.5\columnwidth]{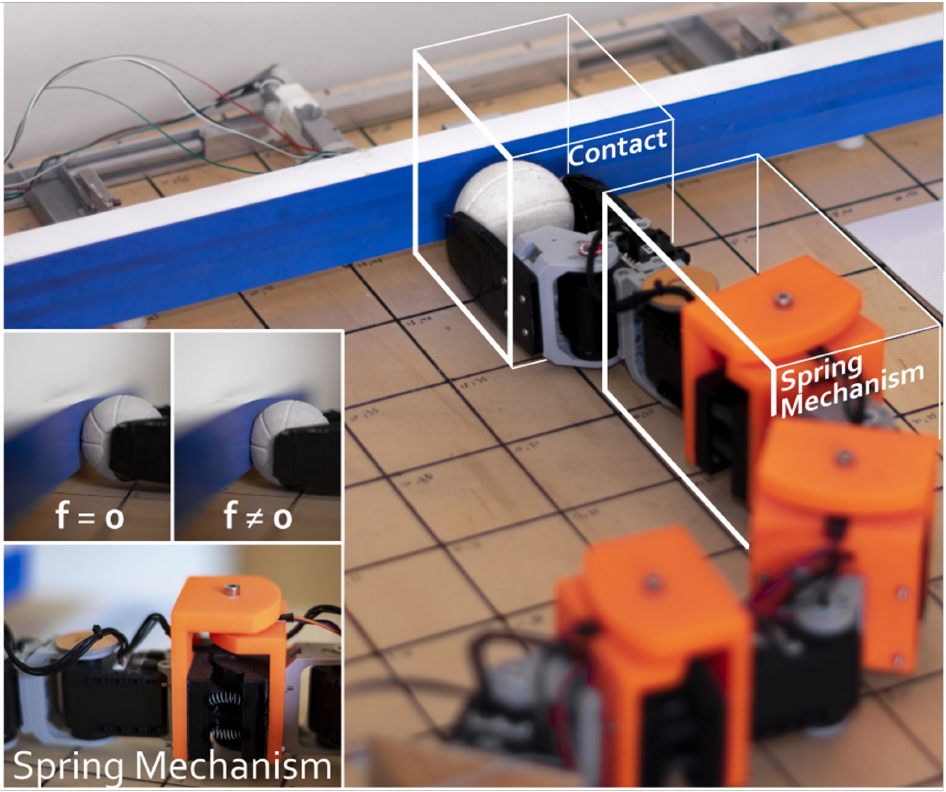}
\caption{Lightweight flexible RM with force capabilities, highlighting the spring mechanism and the contact point.}
\label{fig:photo}
\end{figure}
Moreover, their flexible nature allows the implementation of force/impedance controllers \cite{Colome13, Ficuciello15} and collision detection implementations \cite{DLR2006}, thus becoming a well-known solution for autonomous inspection \cite{sanchez2020fully} and assembly \cite{AEassmebly}. 
%
It is worth mentioning that their advantages in comparison with their rigid counterparts come at the cost of higher sensitivity to control delays, undesired flexible modes and less position accuracy (see e.g. high performance control of rigid-links manipulators in \cite{sanacooll15, AESCTE20}).
\begin{figure*}[t!]
\centering
\includegraphics[width=\linewidth]{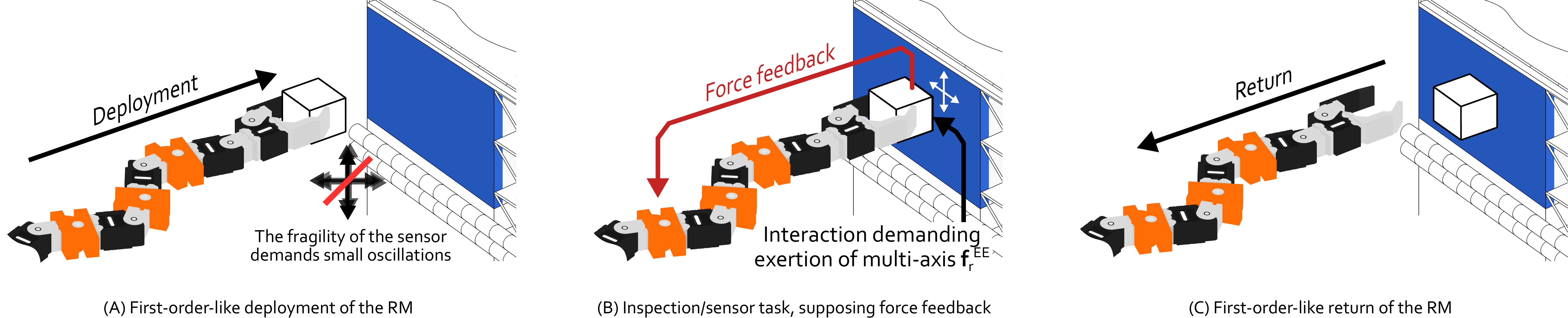}
\caption{Phases of the complex task: inspection/installation of a sensor on an interface with multi-axis force demands.}
\label{fig:mission}
\end{figure*}
To tackle these issues a two-pronged approach is essential: handling parameter uncertainty and rejecting external disturbances.
This is generally pursued either by using i) adaptive techniques, including among others impedance/compliance control \cite{Luo13,siciliano_adaptive}, command-filtered backstepping \cite{Pan2018}, or neural networks \cite{CHENG20092312}; ii) modified Inverse Kinematics (IK henceforth), varying from strategies taking into account the manipulator flexibility \cite{Siciliano2001,Siciliano2000}, to variable  impedance control \cite{Ficuciello15}; or iii) sliding mode control \cite{el2016variable,kashiri2014enhanced}.
Nonetheless, these approaches tend to either compartmentalise the different control modes associated to separate phases of the complete task or depend on estimations with high sensitivity to model uncertainties.
%
In fact, from our own experience, most of contact inspection tasks in non-structured environments are performed by breaking them down into simpler navigation and interaction sub-tasks
(see sketch in Fig. \ref{fig:mission}) and awaiting for stabilisation in each one.

On the other hand, it is well known that the physical interaction changes the whole dynamics thus making manipulation fundamentally a nonlinear problem \cite{hogan1987stable}, often characterised through the mobility tensor.
The proposed control strategy acts like an adaptive impedance interacting with the physical-environment admittance, fulfilling thus the complementarity of physical systems (input=rate and output=force), as shown in the seminal works \cite{hogan1984impedance, hogan1985asme}. Moreover, its adaptive nature makes the controller perform well in both opposite extremes: rigid (low admittance) and soft (high admittance) environments.
Thus, it is a non-trivial extension of the position-only controller of \cite{crdecos_RAL}, where we had to increase its complexity to include force control which is strongly necessary for physical-interaction tasks. Here, the control objectives are the regulation of both position and force \emph{autonomously}, unlike in there. By \emph{autonomously} we mean that the controller itself decides how to adapt its impedance depending on the environmental admittance, to ensure asymptotic regulation to the prescribed references. The approach falls within the framework of closed-loop inverse kinematics, that fits perfectly with the physics of contact and combine position and force feedback thus guaranteeing good performance and stability. Additionally, it regulates the required force vector, not just in normal direction to the contact surface. Overall, the proposed controller for flexible robot manipulators is well suited for complex tasks such as contact inspection, sensor installation and polishing/cleaning.

The contributions are enumerated succinctly below:
\begin{itemize}
\item[C1.] An impedance-like nonlinear adaptive control strategy capable of self-adjusting in mixed contact/non-contact scenarios, i.e. those with variable physical environment admittance. To the best of the authors' knowledge, this is the first IK adaptive  force and position \emph{smooth} controller for flexible robot manipulators that works in unknown environments, rigid and soft.

\item[C2.] Force vector control at the contact point, decomposed in normal and tangential directions.

\item[C3.] Achieving C1 and C2 with a robust controller experimentally validated in a basic microcontroller, demonstrating very-low computational demand and, more importantly, with inexpensive force sensors (Fig. \ref{fig:photo}).
\end{itemize}

The paper is structured as follows: Section~2 outlines the background and framework; Section~3 is devoted to the controller design and main stability results; Section~4 to a thorough experimental validation; and finally, the paper is wrapped up with some conclusions.

\noindent \textbf{Notation:} All vectors are in bold, with $\mathbf{0}_{n}\in \rea{n}$ the zero vector. $0_{n \times m} \in \rea{n \times m}$ denote the zero matrix and $I_n$ the identity. For a matrix $A$, ${\rm ker}(A)$ denotes its kernel, ${\rm rank}(A)$ its rank and if $A>0$, $|\mathbf{v}|_{A} := \mathbf{v}^{\top} A \mathbf{v}$ stands for the vector weighted norm. For a vector $\zeta$,  $J_{\zeta}=\frac{\partial}{\partial \zeta}$ denotes the Jacobian and $J_{\zeta,\nu}$ the sub-block of $J_{\zeta}$ w.r.t the $\nu$-partition of $\zeta$; $\hat \pi$ reads as the $\pi$ estimate, and the operators \emph{column}
\textnormal{col}$(\cdot)$ and \emph{block-diagonal} matrix $\textnormal{diag}(\cdot)$.
{Acronyms:} RM, Robot Manipulator; DoF, Degrees of Freedom; EE, End-Effector; and IK, Inverse Kinematics.


\section{Background and framework}

\begin{figure}[htbp]
\centering
\includegraphics[width=0.8\columnwidth]{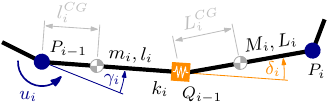}
\caption{Definition of the DoF of the RM.}
\label{fig:dof}
\end{figure}

Let us consider a flexible robot manipulator whose \vspace{-1pt} joint-space configuration reads $\boldsymbol{\theta}:=\textnormal{col}(\boldsymbol{\gamma} , \boldsymbol{\delta})$, where $\boldsymbol{\gamma} \in \rea{N}$ corresponds to the actuated angles and $\boldsymbol{\delta} \in \rea{M}$ to the flexible link deflections, as depicted in Fig. \ref{fig:dof}.
We then define its operational space as $\mathbf{q} := \textnormal{col}({\mathbf{p}},{\boldsymbol{\alpha}}) \in \rea{S}$, where ${\mathbf{p}} \in \rea{S_p}$ denotes the EE position and ${\boldsymbol{\alpha}} \in \rea{S_{\alpha}}$ its orientation, with $S = S_p + S_{\alpha}$.
Thus, the manipulator kinematics reads
\begin{subequations}
\begin{align}
    \dotf{q} &= J \dotfs{\theta} = J_\gamma \dotfs{\gamma} + J_{\delta} \dotfs{\delta} , \label{eq:qdot} \\
    \dotf{p} &= J_p \dotfs{\theta} = J_{p,\gamma} \dotfs{\gamma} + J_{p,{\delta}} \dotfs{\delta} , \label{eq:pdot} \\
    \dotfs{\alpha} &= J_{\alpha} \dotfs{\theta} = J_{\alpha,\gamma} \dotfs{\gamma} + J_{\alpha,{\delta}} \dotfs{\delta} \label{eq:alphadot}.
\end{align}
\end{subequations}
Correspondingly, the joint-space dynamics of the flexible manipulator \cite{de2016robots} for a EE contact force $\mathbf{f} \in \rea{S_p}$ becomes
\begin{equation}
B(\boldsymbol{\theta}) \ddotfs{\theta} + C(\boldsymbol{\theta}, \dotfs{\theta}) \dotfs{\theta} + \mathbf{k}(\boldsymbol{\delta}) + \mathbf{g}(\boldsymbol{\theta}) = \mathbf{u} - J_p^{\top} \mathbf{f} , \label{eq:dyn}
\end{equation}
where $\hspace{-1pt}B,C\hspace{-1pt}$ are the inertia and Coriolis terms matrices; \linebreak $\mathbf{u} := \textnormal{col}(\boldsymbol{\tau},\mathbf{0}_{M})$ includes the control torques $\boldsymbol{\tau} \in \mathbb{R}^N$, $\mathbf{g} := \textnormal{col}(\mathbf{g}_{\gamma},\mathbf{g}_{\delta})$ the gravity, and 
$
\mathbf{k} := \textnormal{col}(\mathbf{0}_N , K \boldsymbol{\delta}) 
$
the flexible torques, with $K \in \mathbb{R}^{M \times M}$ the stiffness of the flexible links \cite{Sicetal09,de2016robots}.


\subsection{Force and motion modeling}
\begin{figure}[htbp!]
\centering
\includegraphics[trim = 0pt 0pt 0pt 22pt , clip , width=0.7\columnwidth]{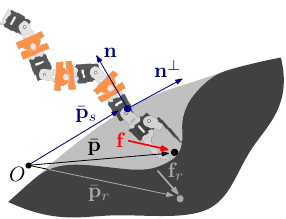}
\caption{Definition of the contact parameters.}
\label{fig:contact}
\end{figure}

Let first define the task-space.
In inverse kinematics, the task-space chosen $\mathbf{s}$ --also denoted as $\mathbf{q}$ in \cite{GALICKI2016165}--, is derived with respect to the actuated DoF and proven to converge to a reference using Lyapunov methods. 
We opt to include the contact forces in our task-space \textit{directly}\footnote{As in \cite{LI2008776}, in contrast to indirect methods, such as \cite{HECK2016235}.} defined as $\mathbf{s} := \textnormal{col}(\mathbf{q} , \mathbf{f})$ and estimate its derivative as
\begin{equation*}
    \dotf{s} = \underbrace{\begin{pmatrix} J_\gamma \dotfs{\gamma} + J_{\delta} \dot{\hat{\boldsymbol{\delta}}} \\ \dot{\hat{\mathbf{f}}} \end{pmatrix}}_{\mathrm{Estimate}} + \underbrace{\begin{pmatrix} J_{\delta} \dot{\tilde{\boldsymbol{\delta}}} \\ \dot{\tilde{\mathbf{f}}} \end{pmatrix}}_{\mathrm{Uncertain}}. \label{eq:goal}
\end{equation*}
Intuitively, we need a model of $\boldsymbol{\delta}, \mathbf{f}$ to \vspace{-1pt} capture the nature of $\dotfs{\delta}, \dotf{f}$. As the focus of this work is on contact, we choose the simple --but accurate-- elastic contact model below.
%
%
For, let first introduce the well-known elastic normal force model (without lateral force), as in \cite{crdecos_RAL}, reading
\begin{equation*}
{\mathbf{f}}_{\top} = {k}_e^{\top} \mathcal{N}^{\top} \left({\mathbf{p}} - {\mathbf{p}}_s \right) ,
\end{equation*}
with $\mathcal{N}^{\top} := \mathbf{n}\mathbf{n}^{\top}$ the projection matrix on the outward contact normal $\mathbf{n} \in \mathbb{R}^{S_p}$, as in Fig. \ref{fig:contact}; $k_e^{\top} \in \mathbb{R^+}$ the \textit{unknown} elastic modulus of the interface; and ${\mathbf{p}}_s \in \mathbb{R}^{S_p}$ the contact point at rest, i.e. without deformation.
This formulation is enriched with elastic lateral forces where $\mathcal{N}^{\perp} := I_{S_p} - \mathcal{N}^{\top}$ denotes the orthogonal projector matrix of $\mathcal{N}^{\top}$ and $k_e^{\perp} \in \mathbb{R^+}$ its \textit{unknown} elastic parameter. 
Defining the elastic vector ${\mathbf{k}}_{e}:=\textnormal{col}({k}_e^{\top} , {k}_e^{\perp})$ and using the matrix Kronecker product $\otimes$
the model reads
%
\begin{equation}
{\mathbf{f}}
= \big({\mathbf{k}}_e^{\top} \otimes I_{S_p} \big) \left(\begin{array}{c} \mathcal{N}^{\top} \\ \mathcal{N}^{\perp} \end{array}\right) \left({\mathbf{p}} - {\mathbf{p}}_s \right)
=: {K}_e \left({\mathbf{p}} - {\mathbf{p}}_s \right). \label{eq:forcemodel}
\end{equation}
If we assume small flexible deflections, then their impact on the contact force is negligible and its derivative reads
\begin{equation}
    \dot{{\mathbf{f}}} = {K}_e J_{p,\gamma} \dotfs{\gamma}. \label{eq:festimation}
\end{equation}
Then, we proceed with the estimation of $\dot{\boldsymbol{\delta}}$. We opt for the inherently pseudo-static formulation in \cite{Siciliano2001} coming from the components in $\boldsymbol{\delta}$ of (\ref{eq:dyn}), but including the proposed and complete force contact model in (\ref{eq:forcemodel}), namely
\begin{equation}
K {\boldsymbol{\delta}} = -J_{p,\delta}^{\top} {\mathbf{f}} +  \mathbf{g}_{\delta},
\label{eq:elasticmodel}
\end{equation}
where $\mathbf{g}_{\delta} := J_{CG,\delta}^{\top} m \mathbf{g}_0$ --with $J_{CG,\delta} \in \mathbb{R}^{S_p \times M}$ linking the centre of mass of the manipulator and the flexible DoF--, $m$ is the mass of the RM, and $\mathbf{g}_0$ is the gravity acceleration in the inertial frame.
Importantly, the factor structure of the proposed contact model \eqref{eq:elasticmodel} is instrumental to deal with the uncertainty of $\mathbf{k}_e$. Moreover, from \eqref{eq:elasticmodel} and using the properties of the Kronecker product an estimate of the flexible deflection rates becomes
\begin{equation}
\dot{{\boldsymbol{\delta}}} = 
{K}^{-1} \Big( (\mathbf{k}_e^{\top} 1) \otimes I_{M} \Big) J_{fg} \dotfs{\gamma}
=: - {\Theta}^{\top} J_{fg} \dotfs{\gamma},  \label{eq:deltadot}
\end{equation}
with the unknown matrix ${\Theta} \in \mathbb{R}^{3M \times M}$
%
%
and the compound force/gravity Jacobian $J_{fg} \in \mathbb{R}^{3M \times N}$
given by
\begin{equation*}
J_{fg} {=} \begin{pmatrix}
\big(I_{2} {\otimes} \dfrac{\partial J_{p,\delta}^{\top}}{\partial \boldsymbol{\gamma}}\big)
\begin{pmatrix} \bar N^{\top} \\ \bar N^{\perp} \end{pmatrix}
+
\big(I_{2} {\otimes} J_{p,\delta}^{\top}\big) \begin{pmatrix} \mathcal{N}^{\top} \\ \mathcal{N}^{\perp} \end{pmatrix} J_{p,\gamma}
\\
- \begin{pmatrix}
\dfrac{\partial J_{CG,\delta}^{\top}}{\partial \gamma_1} & \cdots & \dfrac{\partial J_{CG,\delta}^{\top}}{\partial \gamma_N}
\end{pmatrix} m \bar G
\end{pmatrix},
\end{equation*}
with $\bar N^{\dagger}$, $\bar G \in \mathbb{R}^{N  S_p \times N}$, $\scriptstyle \dagger:= \{\top,\perp\}$, defined column-wise as
\begin{align*}
\bar{\mathbf{N}}^{\dagger}_k &:= \textnormal{col}_{k}\left(
\mathbf{0}_{(k-1) S_p} , (\mathcal{N}^{\dagger} ({\mathbf{p}} - {\mathbf{p}}_s)) , \mathbf{0}_{(N-k) S_p}\right)
, \\
\bar{\mathbf{G}}_k &:= \textnormal{col}_{k}\left(
\mathbf{0}_{(k-1) S_p} , \mathbf{g}_0 , \mathbf{0}_{(N-k) S_p}\right).
\end{align*}
%
%
The model derived above allows to obtain the position and force uncertain error kinematics, where we use the estimates of the unknown elasticity matrices $K_{e}$ and $\Theta$ instead.
Let the position and force errors be defined as $\mathbf{e} := \mathbf{q}_r - \mathbf{q} \in \mathbb{R}^{S}$ and $\boldsymbol{\eta} := \mathbf{f}_r - \mathbf{f} \in \mathbb{R}^{S_p}$, respectively, that are consistent with the task space. Then, using  (\ref{eq:qdot}), (\ref{eq:festimation}) and (\ref{eq:deltadot}) the position and force error kinematics reads
\smallskip
\begin{equation}
\left.\begin{array}{r}
 \\ 
\mathrm{Position:} \\
\mathrm{Force:}
\end{array}\right.
 \! \! \!
\left.\begin{array}{c} \\ \dotf{e} = \\ \dotfs{\eta} = \end{array}\right.
\tcbhighmath[boxrule=0pt,arc=0pt,colback=blue!15!white,top=0pt,bottom=0pt,left=0pt,right=0pt]{
\left.\begin{array}{c}
\mathrm{Estimate} \\
\dotf{q}_r - \hat J_T(\hat \Theta) \dotfs{\gamma} \\
- \hat{K}_e J_{p,\gamma} \dotfs{\gamma}
\end{array}\right.}
\left.\begin{array}{c} \\ {+} \\ {-} \end{array}\right.
\tcbhighmath[boxrule=0pt,arc=0pt,colback=red!15!white,top=0pt,bottom=0pt,left=0pt,right=0pt]{
\left.\begin{array}{c}
\mathrm{Uncertain} \\ 
J_{\delta} \tilde \Theta^{\top} J_{fg} \dotfs{\gamma} \\
\tilde{K}_e J_{p,\gamma} \dotfs{\gamma} 
\end{array}\right.} \label{eq:errordot}
\end{equation}

with $\hat J_T := J_{\gamma} - J_{\delta} \hat \Theta^{\top} J_{fg} \in \mathbb{R}^{S \times N}$ the estimated Jacobian that includes the flexible behaviour from (\ref{eq:deltadot}).
\begin{rem}
In the position-only regulation strategy of \cite{crdecos_RAL} $\dot{\mathbf{q}}_r = \mathbf{0}$, whereas when including the force estimation in the control strategy the evolution of $\dot{\mathbf{q}}_r \neq \mathbf{0}$ 
is used during contact to achieve force regulation $\mathbf{f}=\mathbf{f}_r$, $\dot{\mathbf{f}}_r=0$. 
\end{rem}
%


\section{Unified control strategy}

\begin{figure}[htbp!]
\centering
\includegraphics[width=\columnwidth]{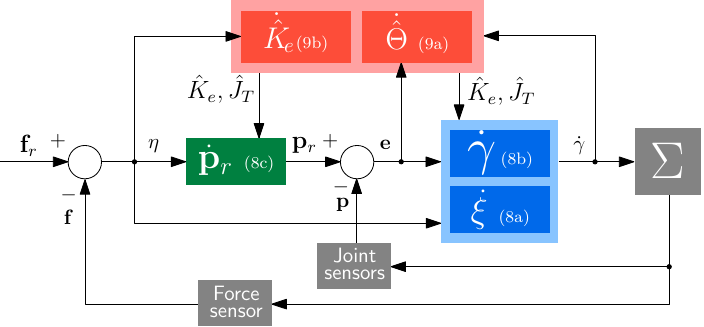}
\caption{Control scheme: integral IK (blue, inner position loop); force-oriented change of references (green, outer force loop); and adaptive laws (orange). Equations in brackets.}
\label{fig:control_scheme}
\end{figure}

The developed model \eqref{eq:errordot} allows to define the adaptive integral IK and nonlinear force controller. For this non-trivial design, we start from the core position-regulation controller of \cite{crdecos_RAL} and augment it with the necessary structure. This includes an integral action $\boldsymbol{\xi} \in \mathbb{R}^S$ motivated from previous work with rigid manipulators of \cite{sanacooll15}. Its final intricate structure is depicted in Fig.~\ref{fig:control_scheme} and reads
\begin{subequations}
\begin{align}
\dotfs{\xi} &= - K_{\xi} \boldsymbol{\xi} + K_I \hat J_T K_{\gamma} ( \hat J_T^{\top} K_P \mathbf{e} + J_{p,\gamma}^{\top} \hat{K}_e \boldsymbol{\eta} ) , \label{eq:control_xi}
\\
\dotfs{\gamma} &=  K_{\gamma} \hat J_T^{\top} \left( K_P \mathbf{e} + K_I\boldsymbol{\xi} \right) + K_{\eta} J_{p,\gamma}^{\top} \hat{K}_e \boldsymbol{\eta}, \label{eq:control_gamma}
\\
\dotf{p}_r &=  \hat J_T K_{\gamma \eta} J_{p,\gamma}^{\top} \hat{K}_e \boldsymbol{\eta}
- \sigma(\normv{\boldsymbol{\eta}}) K_P \mathbf{e},  \label{eq:control_pr}
\end{align}
\label{eq:control_core}
\end{subequations}
where the key scalar function $\sigma(\normv{\boldsymbol{\eta}})$, $\sigma(0)=0$, is positive definite; together with the adaptive update laws to cope with the uncertain contact model given by 
\begin{subequations}
\begin{align}
\dot{\hat \Theta} &= \Gamma_{\Theta} J_{fg} \dotfs{\gamma} \hspace{0.03cm} \mathbf{e}^{\top} K_P J_{\delta}, \quad \hat \Theta(0)= \Theta_{0}, \label{eq:thetahatdot}
\\
\dot{\hat{k}}_e^{\dagger} &= \mathrm{Proj}( {\varpi}_{\mathcal{N}^{\dagger}}, \rho({\hat{k}}_e^{\dagger})), \quad \rho(\hat{k}_e^{\dagger}(0)) \leq 1, \label{eq:keperphatdot}
\end{align}
\label{eq:adaptive_laws}
\end{subequations}
where
%
recall that we define the upper-index symbols \mbox{$\scriptstyle \dagger= \{\top,\perp\}$} as the perpendicular projections of the force vector, 
and the projector operator $\mathrm{Proj}(\cdot,\cdot)$ \cite{Krsticbook, IoaSun} associated to an suitable convex scalar function $\rho$;
and 
%
\begin{align}
{\varpi}_{\mathcal{N}^{\dagger}} (\boldsymbol{\xi}, {\mathbf{e}}, \boldsymbol{\eta}) :=& \ - \Gamma_{\mathcal{N}^{\dagger}} \ \boldsymbol{\eta}^{\top} {\mathcal{N}^{\dagger}} J_{p,\gamma} \dotfs{\gamma}, \label{eq:wn}
\end{align}
%
being $K_{P}, K_{I}, K_\xi, K_{\gamma}, K_{\eta}$ and $\Gamma_{\Theta}$ positive definite matrices of appropriate dimensions, and $\Gamma_{\mathcal{N}^{\dagger}}\in \rea{+}$.
\begin{rem}
We underscore that 
the stiffness $K_{e}$ is completely unknown and the adaptation is in several force directions. Moreover, as it will be shown further, this controller is able to decide when to use the force and/or position depending on the actual physical interaction.
Finally, even though the control architecture shows inner/outer-like position and  force loops they are not in cascade but being strongly coupled with the adaptive update laws and the state interconnection through $(\ref{eq:control_pr})$ (see Fig.~\ref{fig:control_scheme}). Nonetheless, we show in the experiments that it can be implemented in a basic microcontroller.
\end{rem}

\subsection{Main stability result}

Before stating the main stability result, let us rearrange the closed-loop kinematics 
plugging (\ref{eq:control_core}) in (\ref{eq:errordot}) and defining the state vector $\mathbf{x} := \textnormal{col}(\boldsymbol{\xi}, \mathbf{e}, \boldsymbol{\eta})$. These result in
\begin{equation}
\dotf{x}
= \hat{\mathcal{A}} (\mathbf{q}, \hat \Theta, \hat{K}_e) \mathbf{x} + \tilde{\mathcal{B}} (\mathbf{q}, \tilde \Theta, \tilde{K}_e) \dotfs{\gamma}, \label{eq:closedloop}
\end{equation}
with the estimation matrix $\hat{\mathcal{A}}$ defined as
\begin{align*}
\hat{\mathcal{A}} 
&:=
\left(\begin{array}{cc:c}
- K_{\xi} & K_I \hat{\mathcal{J}}_T K_P & K_I \hat{\mathcal{J}}_{T \gamma} \hat{K}_e
\\
- \hat{\mathcal{J}}_T K_I & - \big( \hat{\mathcal{J}}_T + \sigma (\normv{\boldsymbol{\eta}}) I_{S} \big) K_P & \hat{\mathcal{J}}_{T \gamma} \hat{K}_e
\\ \hdashline
-\hat{K}_e \hat{\mathcal{J}}_{T \gamma}^{\top} K_I & -\hat{K}_e \hat{\mathcal{J}}_{T \gamma}^{\top} & - \hat{K}_e \mathcal{J}_{\gamma} \hat{K}_e
\end{array}\right), 
\end{align*}
denoting $\hat{\mathcal{J}}_T := \hat J_T K_{\gamma} \hat J_T^{\top}$, $\mathcal{J}_{\gamma} := J_{p,\gamma} K_{\eta} J_{p,\gamma}^{\top}$, $\hat{\mathcal{J}}_{T \gamma} := \hat J_T K_{\gamma} J_{p,\gamma}^{\top}$ and
$$
\tilde{\mathcal{B}}:=\left(\begin{array}{cc:c} 0_{N \times S} &  J_{fg}^{\top} \tilde \Theta J_{\delta}^{\top} & - J_{p,\gamma}^{\top}\tilde{K}_e^{\top} \end{array}\right)^{\top}.$$


%
The next proposition states the main stability result.

\begin{thm}[Force control] \label{prop:main}
Consider the uncertain error system (\ref{eq:errordot}) 
and assume $\mathrm{rank}[J_q] = S$ and $\mathrm{rank}[J_{p,\gamma}] = S_p$ together with the design conditions $N\geq S$ and $N \geq S_{p}$. Define the function $\rho$ of $\mathrm{Proj}(\cdot,\rho)$ in \eqref{eq:keperphatdot} convex and such that $ | \hat{k}_e^{\dagger}(t) | > \epsilon$ for some constant $\epsilon > 0$, $t \geq 0$; and $\sigma(\normv{\boldsymbol{\eta}})$, $\sigma(0)=0$ positive definite.
Then, for any reference force $\mathbf{f}_r$ the adaptive-state feedback given by \eqref{eq:control_core} and \eqref{eq:adaptive_laws}-\eqref{eq:wn} guarantees
global boundedness of all signals $\mathbf{x}, \mathbf{p}_r,  {\hat \Theta}$ and 
$\hat{\mathbf{k}}_{e}$ and (globally) asymptotically stable zero equilibrium $\mathbf{x}=\mathbf{0}$.
\end{thm}
\begin{proof}
Let us define the radially unbounded and positive definite Lyapunov function candidate given by
\begin{equation}
V := \dfrac{1}{2} \normv{\mathbf{x}}^2_{\mathcal{K}} + \dfrac{1}{2} \mathrm{Tr} \left( \tilde{\Theta}^{\top} \Gamma_{\Theta}^{-1} \tilde{\Theta} \right) 
+ \dfrac{1}{2} |{\tilde{\mathbf{k}}_{e}}|^2_{\Gamma^{-1}_{e}},
\end{equation}
with $\mathcal{K} :=\textnormal{diag}(I_S,K_P,I_{S_p})$ and ${\Gamma_e} :=\textnormal{diag}(\Gamma_{k_e^{\top}},\Gamma_{k_e^{\perp}})$.
Its derivative along the trajectories of (\ref{eq:closedloop}) reads
\begin{equation*}
    \dot V \hspace{-1pt} = \hspace{-0.5pt}
\mathbf{x}^{\top} \mathcal{K} \hat{\mathcal{A}}\mathbf{x} \hspace{-0.5pt}+\hspace{-0.5pt} \mathbf{x}^{\top} \mathcal{K} \tilde{\mathcal{B}} \dotfs{\gamma} \hspace{-0.5pt}+\hspace{-0.5pt} \mathrm{Tr} \left( \tilde \Theta^{\top} \Gamma_{\Theta}^{-1} \dot{\tilde{\Theta}} \right) \hspace{-0.5pt}
+ {\tilde{\mathbf{k}}_{e}}^{\top} {\Gamma^{-1}_{e}} \dot{\tilde{\mathbf{k}}}_{e}.
\end{equation*}
%
%
This expression can be rewritten as
\begin{align*}
\dot V =&  
- \boldsymbol{\xi}^{\top} K_{\xi} \boldsymbol{\xi} 
- \bar{\mathbf{e}}^{\top} ( \hat{\mathcal{J}}_T + \sigma \normv{\boldsymbol{\eta}} ) \bar{\mathbf{e}}
- \boldsymbol{\eta}^{\top} \hat{K}_e \mathcal{J}_{\gamma} \hat{K}_e \boldsymbol{\eta}
\\&+ \bar{\mathbf{e}}^{\top} J_{\delta} \tilde \Theta^{\top} J_{fg} \dotfs{\gamma} - \mathrm{Tr} \left( \tilde \Theta^{\top} \Gamma_{\Theta}^{-1} \dot{\hat{\Theta}} \right) 
\\&- \boldsymbol{\eta}^{\top} \tilde{K}_e J_{\gamma}^t \dotfs{\gamma}   
- {\tilde{\mathbf{k}}_{e}}^{\top} {\Gamma^{-1}_{e}} \dot{\hat{\mathbf{k}}}_{e}
\\
=& 
- \boldsymbol{\xi}^{\top} K_{\xi} \boldsymbol{\xi} 
- \bar{\mathbf{e}}^{\top} ( \hat{\mathcal{J}}_T + \sigma \normv{\boldsymbol{\eta}} ) \bar{\mathbf{e}}
- \boldsymbol{\eta}^{\top} \hat{K}_e \mathcal{J}_{\gamma} \hat{K}_e \boldsymbol{\eta}
\\&+ \mathrm{Tr} \left[ \tilde \Theta^{\top} \left(- \Gamma_{\Theta}^{-1} \dot{\hat{\Theta}} + J_{fg} \dotfs{\gamma} \bar{\mathbf{e}}^{\top} J_{\delta} \right) \right] \\
&- {\tilde{\mathbf{k}}_{e}}^{\top} {\Gamma^{-1}_{e}} 
\left(\begin{array}{c} 
\mathrm{Proj}(\boldsymbol{\varpi}_{\mathcal{N}^{\top}}, \rho({\hat{k}}_e^{\top})) - \boldsymbol{\varpi}_{\mathcal{N}^{\top}}  \\
\mathrm{Proj}( \boldsymbol{\varpi}_{\mathcal{N}^{\perp}}, \rho({\hat{k}}_e^{\perp})) - \boldsymbol{\varpi}_{\mathcal{N}^{\perp}}
\end{array}\right)
\end{align*}
where we have used $\bar{\mathbf{e}} := K_P {\mathbf{e}}$ for compactness as well as the trace properties. 
By definition of the projector operator and the proposed update laws, we get
\begin{equation*}
    \dot V \leq - \boldsymbol{\xi}^{\top} K_{\xi} \boldsymbol{\xi} 
    - \bar{\mathbf{e}}^{\top} ( \hat{\mathcal{J}}_T + \sigma \normv{\boldsymbol{\eta}} ) \bar{\mathbf{e}}
    - \boldsymbol{\eta}^{\top} \hat{K}_e \mathcal{J}_{\gamma} \hat{K}_e \boldsymbol{\eta}. 
\end{equation*}
Therefore, since $V>0$ globally and $\dot V \leq 0$, all the trajectories are bounded and the first claim follows.
%
%

For the stability claim and since the closed-loop is autonomous we analyse the largest invariant set included in the residual dynamics and invoke LaSalle's Invariance Principle. Let first prove that the matrix $\hat{K}_e$ is invertible. By requirement the projectors ensure $\hat{k}_e^{\top}, \hat{k}_e^{\perp} \neq 0$, then
\begin{align*}
    \det (\hat{K}_e) &= \det \left[ \hat{k}_e^{\top} \mathbf{n} \mathbf{n}^{\top} + \hat{k}_e^{\perp} \left( I_{S_p} - \mathbf{n} \mathbf{n}^{\top} \right) \right] 
    \\
    &= (\hat{k}_e^{\perp})^{S_p} \det \left[I_{S_p} + \left({\hat{k}_e^{\top}} / {\hat{k}_e^{\perp}} - 1\right) \mathbf{n} \mathbf{n}^{\top} \right] 
    \\
    &= (\hat{k}_e^{\perp})^{S_p-1} \hat{k}_e^{\top} \neq 0 ,
\end{align*}

where the matrix determinant lemma has been invoked.   
From above we can define the vector $\hat{\boldsymbol{\eta}}:=\hat{K}_e \boldsymbol{\eta}$. Thus, the residual dynamics are defined in the set 
$$\Omega := \left\{ \mathbf{x}\in \mathbb{R}^{2S+S_p};\hat{\Theta} \in \mathbb{R}^{3M \times M};\hat{\mathbf{k}}_{e} \in \mathbb{R}^{2}  : \dot V = 0 \right\}.$$
We distinguish two cases $\boldsymbol{\eta} \neq \mathbf{0}$ and $\boldsymbol{\eta} = \mathbf{0}$, where it is straightforward to check, from \eqref{eq:control_core} and \eqref{eq:closedloop}, that the dynamics restricted to $\Omega$ become \mbox{$\dotfs{\gamma}=\dotf{x}=\dot{\hat \Theta}=\dot{\hat{\mathbf{k}}}_{e}=\mathbf{0}$}, hence containing only fixed points and reducing the analysis to rule out all the undesirable equilibria.

%
[\noindent{Case $\boldsymbol{\eta} \neq \mathbf{0}$}].
Notice that, the function $\sigma$ is positive definite, and hence $\dot V = 0\Leftrightarrow \boldsymbol{\xi}=\mathbf{e}= \mathbf{0}$ and \mbox{$J_{p,\gamma}^{\top} \hat{\boldsymbol{\eta}} = \mathbf{0}$}. The assumption $\mathrm{rank}(J_{p,\gamma}) = S_p$ and the condition $N\geq S_{p}$ imply $\ker\big[J_{p,\gamma}^{\top}\big]=\mathbf{0}$ and therefore inducing injectivity that ensures $\hat{\boldsymbol{\eta}}=\mathbf{0}$.

\smallskip
[\noindent{Case $\boldsymbol{\eta} = \mathbf{0}$}].
In this case, by definition $\sigma_{p}=0$ and hence \mbox{$\dot V = 0\Leftrightarrow \boldsymbol{\xi}=\boldsymbol{\eta}= \mathbf{0}$} and $\hat{J}^{\top}_T \bar{\mathbf{e}}=\mathbf{0}$. 
Notice that $\hat{J}_T^{\top} = (\begin{array}{c:c} I_N & - J_{fg}^{\top} \hat \Theta \end{array} ) J_{q}^{\top}$ and let us define $v:=J_{q}^{\top} \bar{\mathbf{e}}\in \mathbb{R}^{N+M}$. By assumptions $\mathrm{rank}(J_q) = S$ and $N\geq S$ thus inducing and injective map from $\bar{\mathbf{e}}$ to $v$. Hence,  \mbox{$\hat{J}^{\top}_T \bar{\mathbf{e}}=\mathbf{0}$} implies that $v \in \ker\left[ (\begin{array}{c:c} I_N & - J_{fg}^{\top} \hat \Theta \end{array}) \right]$ that is spanned by the columns of
$J_{\hat \Theta}^{\perp} := (\begin{array}{c:c} \hat \Theta^{\top} J_{fg} & I_M \end{array} )^{\top}$
and then $v=J_{\hat \Theta}^{\perp} \varrho$ for some arbitrary $\varrho \in \mathbb{R}^{M}$. Consider now the linear map $v=J_{\hat \Theta}^{\perp} \varrho$ which has $\mathrm{rank}\big[J_{\hat \Theta}^{\perp} \big] = M$. Thus, $\ker\big[J_{\hat \Theta}^{\perp} \big] = \mathbf{0}$ which implies $J_{\hat \Theta}^{\perp} \varrho = \mathbf{0} \Leftrightarrow \varrho = \mathbf{0}$ and therefore $v = \mathbf{0} \Rightarrow \mathbf{e} = \mathbf{0}$.
\qed
\end{proof}

\begin{cor}[Unifying property] \label{th:cor}
Consider a non-contact case $\boldsymbol{\eta} \neq \mathbf{0}$ with a constant reference position $\mathbf{p}_r=\mathbf{p}_r^{*}$, with $\dot{\mathbf{p}}_r=0$ from \eqref{eq:control_pr}. 
Then, the adaptive-state feedback  \eqref{eq:control_xi}-\eqref{eq:control_gamma} and \eqref{eq:adaptive_laws}-\eqref{eq:wn} guarantees (globally) asymptotically stable equilibrium $\mathbf{p}=\mathbf{p}_r^{*}$, $\dot{\boldsymbol{\eta}} = \mathbf{0}$.
\end{cor}
\begin{proof}
This result is a direct consequence from the case $\boldsymbol{\eta} \neq \mathbf{0}$ in the proof of Theorem \ref{prop:main}. \qed
\end{proof}
\begin{rem}
Although straightforward from Theorem \ref{prop:main}, Corollary \ref{th:cor} has been stated to highlight the unifying property. This means that the controller is able to follow alternate position and force demands without the need of changing its  architecture, and hence being able to accomplish complex tasks as the one described in Fig. \ref{fig:mission}. To demosntrate this fact specifically, we propose a hybrid position-force scenario in the experimental validation.
\end{rem}


\section{Experimental validation results}

\begin{figure}[htbp]
\centering
\includegraphics[trim = 80pt 0pt 0pt 10pt , clip , width=\columnwidth]{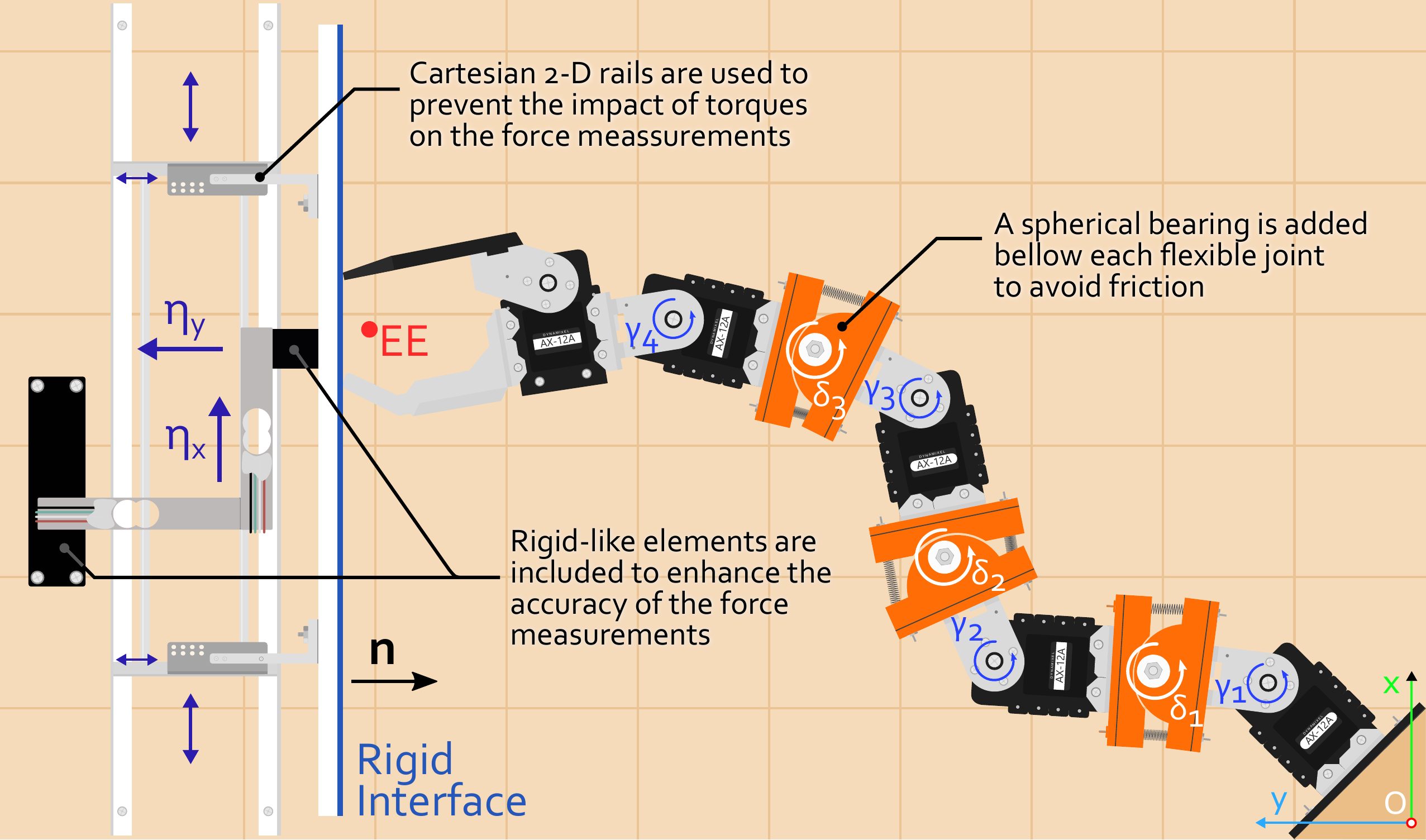}
\caption{Experimental setting including rigid contact interface and force measurement system. Friction can be neglected due to the spherical bearings in the flexible joints.}
\label{fig:exp_setting}
\end{figure}

To demonstrate the motion/force control capabilities, an experimental validation scenario is proposed: the tracking of a set of waypoints and multi-axis forces on a contact interface equipped with cheap force sensors shown in Fig.~\ref{fig:exp_setting}.
The robot manipulator is composed of $N=4$ actuated DoF mounting Dynamixel AX-12A servomotors --with a resolution of 0.3\si{\degree} (0.0052 \si{\radian}), a stall torque of 1.5 \si{\newton \meter} and lacking speed control mode
-- and $M=3$ flexible joints.
%
%
The latter are designed with a low stiffness dual spring system --whose deflection is measured with Murata SV01 potentiometers--, being prone to nonlinear behaviours and, hence, highlighting the robustness of the control strategy.
These two link typologies are placed alternatively, actuated followed by flexible (such as in Fig. \ref{fig:dof}), except for the last DoF that lacks the flexible one. All the parameters are shown in Table \ref{tab:param}. 

\bigskip
\begin{table}[htbp]
\caption{Benchmark application parameters}
\centering
\begin{scriptsize}
\begin{tabular}{@{}cccccccccc@{}}
\multicolumn{
7}{c}{Compound joints [\si{\centi\meter}, \si{\gram}, \si{\newton \meter \per \radian}]} & \multicolumn{
3}{c}{EE joint [\si{\centi\meter}, \si{\gram}]} \\ \cmidrule(r){1-7} \cmidrule(l){8-10}
$l$ & 
$L$ & 
$l^{CG}$ & 
$L^{CG}$ & 
$m$ & 
$M$ & 
$k$ & 
$l_{EE}$ & 
$l^{CG}_{EE}$ & 
$m_{EE}$ \\ \midrule
4.8 &  
6.2 &  
2.4 &  
3.6 & 
25 &  
64 &  
0.8 & 
12.0 & 
6.0 & 
72 
\end{tabular}
\end{scriptsize}
\label{tab:param}
\end{table}
\begin{table*}[t!]
\caption{Benchmark control gains}
\centering
\begin{scriptsize}
\begin{tabular}{@{}cccccccccccccc@{}}
\multicolumn{2}{c}{$K_P$} & \multicolumn{2}{c}{$K_I$} & \multicolumn{2}{c}{$K_{\xi}$} & \multicolumn{4}{c}{$K_{\gamma}$} & \multicolumn{4}{c}{$K_{\eta}$} \\ 
\cmidrule(r){1-2} \cmidrule(lr){3-4} \cmidrule(lr){5-6} \cmidrule(lr){7-10} \cmidrule(lr){11-14}
$\bar{\mathbf{p}}$ & 
$\bar{\boldsymbol{\alpha}}$ & 
$\bar{\mathbf{p}}$ & 
$\bar{\boldsymbol{\alpha}}$ &  
$\bar{\mathbf{p}}$ & 
$\bar{\boldsymbol{\alpha}}$ & 
$1$ & $2$ & $3$ & $4$ & $1$ & $2$ & $3$ & $4$  \\ \midrule
0.5949 & 0.0214 & 0.1610 & 0.0024 & 0.1200 & 0.1200 & 209.9 & 220.5 & 241.4 & 283.4 & 20.00 & 20.00 & 20.00 & 20.00  
\\ \vspace{-10pt}
\\ 
\multicolumn{6}{c}{$\Gamma_{\Theta}$} \\ 
\cmidrule(r){1-6} 
$\Theta_1$ & $\Theta_2$ & $\Theta_3$ & $\Theta_4$ & $\Theta_5$ & $\Theta_6$ & $\sigma_{p}$ & $\Gamma_{k_e^{\top}}$ & $\Gamma_{k_e^{\perp}}$ & $k_M^{\top}$ & $k_m^{\top}$ & $k_M^{\perp}$ & $k_m^{\perp}$ & $\beta$ \\ \midrule
257.1 & 1929 & 3857 & 51.43 & 385.7 & 771.4 & 0.3 & 0.0040 & 0.0020 & 0.0120 & 0.0040 & 0.0120 & 0.0040 & 0.4000
\end{tabular}
\end{scriptsize}
\label{tab:gains}
\end{table*}

On the other hand, the dual-axis force sensor is formed by joining two linear bi-directional load cells in a $\Gamma$-shape as depicted in Fig.~\ref{fig:exp_setting} and converting their signals using an HX711 ADC, thus obtaining relatively clean force measurement. Altogether, the system provides enough information to close the loop with acceptable accuracy. 

To compute the feedback and the control algorithm, an Arduino Mega 2560, based on a 16 \si{\mega\hertz} ATmega2560 and with a flash memory of 256 \si{\kibi\byte} and an SRAM of 8 \si{\kibi\byte}, is used. This controller board provides a stable frequency of 40 \si{\hertz} throughout the experiment. We have chosen a planar RM for the validation just to simplify the force-measurement system and alleviate the computation load for this cheap microcontroller, but keeping the novelty of the vector force compensation.
%
Finally, a projector operator consistent with Theorem \ref{prop:main} yields
\begin{equation*}
\mathrm{Proj}({\varpi} , \rho({\hat\kappa})) := \left\lbrace
\begin{array}{ll} 
\left[ 1-(\rho(\hat\kappa)')^{2} \right]\rho(\hat\kappa) & {\varpi}, \  \rho(\hat\kappa)' {\varpi} > 0 
\\
{\varpi} & \mathrm{otherwise}
\end{array}
\right. ,
\end{equation*}
being $\hat\kappa \in \mathbb{R}$ the parameter projected and $\rho$ the convex function which has been defined as

\begin{equation*}
\rho({\hat\kappa}) := \dfrac{\left( \hat\kappa - \dfrac{\kappa_M + \kappa_m}{2}\right)^2}{\left( 1-\beta^2 \right)\left(\dfrac{\kappa_M - \kappa_m}{2}\right)^2} - \dfrac{\beta^2}{1-\beta^2}  .
\end{equation*}

The positive definite function $\sigma$ is defined as $\sigma:=\sigma_p \normv{\boldsymbol{\eta}}$, with constant $\sigma_p>0$  and the control gains and parameters used in \emph{all} the experiments are shown in Table~\ref{tab:gains}.



%
With the above experimental settings we propose two scenarios to validate the approach: 
i) an exclusively \textit{ad-hoc} force control setting to validate the result of Theorem \ref{prop:main}; and ii) a mixed scenario including unperturbed displacement and controlled contact forces, to validate the universal property combining the results of Theorem \ref{prop:main} and Corollary \ref{th:cor}, hence providing an example of application with an \textit{implicit} transition between the two modes.
%
We underscore that these experiments also validate the computer implementation of the algorithm in a cheap microcontroller and the hardware control system.

\subsection{Contact force scenario (force control)}

\begin{figure}[htbp!]
\centering
\begin{overpic}[abs,unit=1mm,width=0.8\columnwidth]{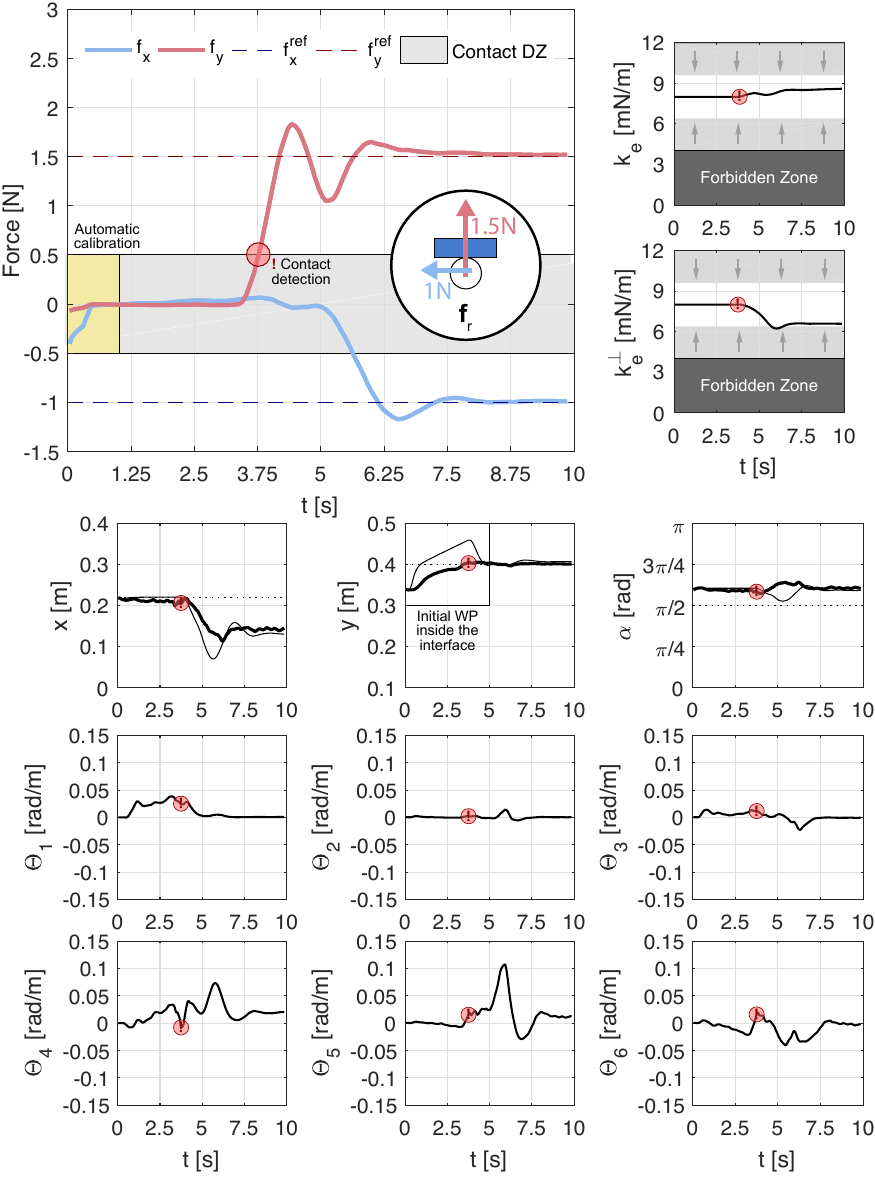} 
\put(84.5,141.5){\rotatebox[origin=c]{90}{\scalebox{0.5}{$\top$}}}
\end{overpic}
\caption{Results of the contact force case: force tracking and adaptive contact parameters (top), with projector influence and contact detection; Cartesian position and orientation (centre, with modified references as lighter lines in contrast to the original ones, dotted); and $\hat \Theta$ parameters (bottom).}
\label{fig:exp2}
\end{figure}

\begin{figure*}[t!]
\centering
\begin{overpic}[abs,unit=1mm,width=0.98\textwidth]{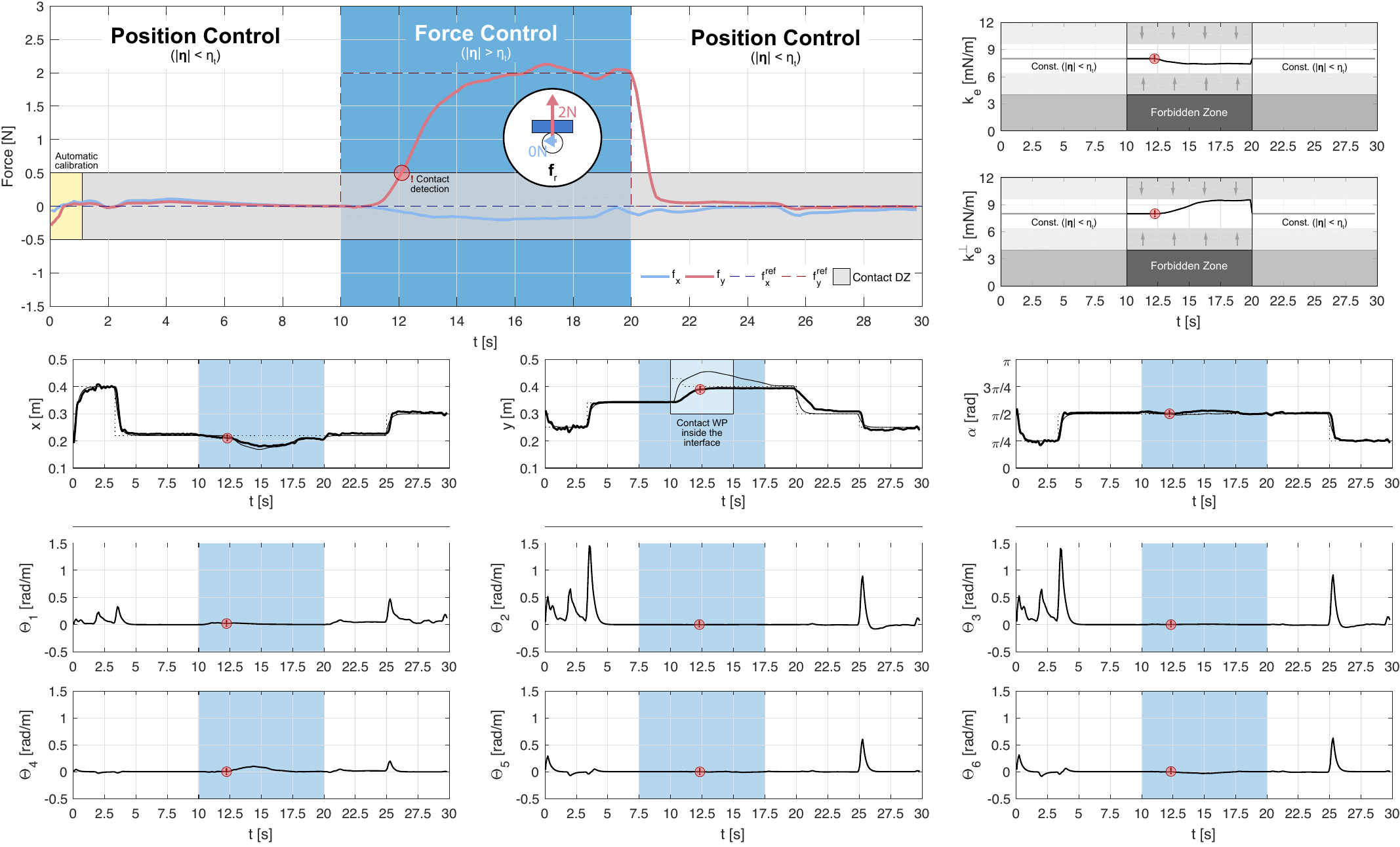} 
\put(101.2,79.1){\rotatebox[origin=c]{90}{\scalebox{0.4}{$\top$}}}
\end{overpic}
\caption{Results of the mixed contact/non-contact scenario, with: force regulation and adaptive contact parameters (top), including the projector influence zones; the Cartesian position and orientation (centre, with modified references as lighter lines and original ones dotted); and the whole set of adaptive parameters (bottom).}
\label{fig:exp3}
\end{figure*}

The force control---second phase in Fig. \ref{fig:mission} B)--, is demonstrated for $\mathbf{f}_r = [-1,  1.5 ]^{\top}$\si{\newton} and an initial condition close to the contact interface. As shown in Fig. \ref{fig:exp2}, these force demands are achieved while the outer position loop derived from the modification of $\dot{\mathbf{p}}_r$ in (\ref{eq:control_pr}) converges to the trivial error. To obtain this outcome, nonetheless, some other parameters have a significant, but indirect, contribution. On the one hand, the adaptive contact parameters $\hat{k}_e^{\top}, \hat{k}_e^{\perp}$ show slight variations that shape the rate of change of the position reference to obtain a safer response: decreasing its sensitivity to smooth the convergence when the forces are bellow their references and increasing it when these are excessive. On the other hand, the adaptive parameters $\hat \Theta$ are responsible of dampening the possible oscillations that could occur just after the first contact.
%
Altogether, for a good validation the main contribution of Theorem \ref{prop:main} is demonstrated in this \textit{ad-hoc} and clean scenario minimising other external factors, except noise and cheap computer implementation.

\subsection{Mixed scenario (unified position and force control)}

The second experiment, and the most interesting one showing the unified property, includes the interaction between the contact force and position reference changes. This scenario imitates the motivational task describen in the Fig. \ref{fig:mission}, where the controller has to follow a position-force-position references corresponding to the three phases.
As it was shown in Corollary \ref{th:cor}, the controller decides autonomously the `soft switching' between them. These soft transitions are achieved without using any control switch or changing the control gains, just by the inherent cancellation of the outer loop when $\boldsymbol{\eta} = \mathbf{0}$. However, in practice the noise and the cheap force sensor hardware preclude the use of the equality condition and hence a threshold $\eta_{t}$ is introduced for its implementation as $\lvert\boldsymbol{\eta}\lvert < \eta_{t} \rightarrow \boldsymbol{\eta} = \mathbf{0}$, with constant $\eta_{t} >0$. The force reference is set as $\mathbf{f}_r = [0,  2 ]^{\top}$\si{\newton}, i.e. pressing without inducing tangential forces.

As thoroughly shown in Fig. \ref{fig:exp3}, the interaction between these has no noticeable impact on the response, showing a first-order-like response when commanded to move before and after the contact. 
Additionally, it is worth going into the details of the indirect parameters, as done with the previous scenario. Firstly, the contact parameters show wider variations in this case (when they are not cancelled by $\lvert\boldsymbol{\eta}\lvert < \eta_{t}$ before and after contact), not having any significant impact on the force tracking convergence apart from the mere experimental variability. Secondly, the results of the matrix of adaptive parameters associated with the flexibility of the RM are completely consistent with the previous scenario. However, two comments ought to be included: i) the aforementioned difference in their order of magnitude between contact and displacement make their variation in the former case indistinguishable; and ii) the terms associated to $\mathcal{N}^{\perp}$ in $\hat \Theta$ are quite noticeable after contact, the only difference with respect to the previous scenario.


\section{Conclusions}
An adaptive force and motion control strategy for flexible manipulators is presented. Thanks to its adaptive nature, this controller is well-suited for the installation of fragile devices demanding controlled force. While its multi-axis force capabilities allow performing contact tasks converging to these force demands, the adaptation to the stiffness of the flexible links provides disturbance rejection capabilities, that increase the accuracy of the operation and reduce the risk of incidents.
Together with a detailed derivation using well-known Lyapunov-based control techniques, full theoretical and experimental results are provided. Among the latter, it is worth highlighting the scenario including both abrupt position reference changes and force regulation, thus demonstrating the suitability of the control strategy for applications with mixed contact/non-contact scenarios that provide  variable environmental admittance. 
%



\section*{Acknowledgement}
This work was supported by the Project HOMPOT grant number P20\_00597 under framework PAIDI 2020.


\bibliographystyle{elsarticle-harv}
\bibliography{RM}              

\end{document}